\documentclass[sigconf]{acmart}
\renewcommand\footnotetextcopyrightpermission[1]{} 
\setcopyright{none}
\AtBeginDocument{%
  \providecommand\BibTeX{{%
    \normalfont B\kern-0.5em{\scshape i\kern-0.25em b}\kern-0.8em\TeX}}}

\setcopyright{acmcopyright}
\copyrightyear{2021}
\acmYear{2021}
\acmDOI{10.1145/1122445.1122456}

\acmConference[SIGIR '21]{ACM SIGIR Conference on Research and Development in Information Retrieval}{July 11--15, 2021}{Online}
\acmBooktitle{ACM SIGIR Conference on Research and Development in Information Retrieval, July 11--15, 2021, Online}
\acmPrice{15.00}
\acmISBN{978-1-4503-XXXX-X/18/06}

\usepackage{multirow}
\usepackage{tabularx}
\usepackage{siunitx}
\usepackage{xcolor}

\newcommand{\set}{\mathbb}

\newcommand{\appname}{QuTI}

\begin{document}

\title{\appname{}! Quantifying Text-Image Consistency in Multimodal Documents} 

\author{Matthias Springstein}
\email{matthias.springstein@tib.eu}
\orcid{1234-5678-9012}
\affiliation{%
  \institution{TIB -- Leibniz Information Centre for Science and Technology}
  \city{Hannover}
  \country{Germany}
}

\author{Eric Müller-Budack}
\email{eric.mueller@tib.eu}
\orcid{1234-5678-9012}
\affiliation{%
  \institution{TIB -- Leibniz Information Centre for Science and Technology}
  \city{Hannover}
  \country{Germany}
}

\author{Ralph Ewerth}
\email{ralph.ewerth@tib.eu}
\orcid{1234-5678-9012}
\affiliation{%
  \institution{TIB -- Leibniz Information Centre for Science and Technology}
  \institution{L3S Research Center, Leibniz University Hannover}
  \city{Hannover}
  \country{Germany}
}

\renewcommand{\shortauthors}{Springstein et al.}

\begin{abstract}
The World Wide Web and social media platforms have become popular sources for news and information. Typically, multimodal information, e.g., image and text is used to convey information more effectively and to attract attention. While in most cases image content is decorative or depicts additional information, it has also been leveraged to spread misinformation and rumors in recent years. In this paper, we present a Web-based demo application that automatically quantifies the cross-modal relations of entities~(persons, locations, and events) in image and text. The applications are manifold. For example, the system can help users to explore multimodal articles more efficiently, 
or can assist human assessors and fact-checking efforts in the verification of the credibility of news stories, tweets, or other multimodal documents. 
\end{abstract}



\keywords{Cross-modal consistency, image-text-relations, deep learning, multimodal documents}


\maketitle

\section{Introduction}
\label{sec:intro}
With the availability of digital environments, the World Wide Web and social media have become popular sources of information~\cite{Rogers2013, Broersma2013, TandocJr2016}. 
However, social media also allows individual users to publish and disseminate paraphrased or manipulated news stories that convey an intended narrative or opinion. As one of the consequences, \textit{fake news}, i.e., articles that deliberately spread rumors or misleading information, have become a critical problem in recent years and have even been used repeatedly during the \textit{2016 United States elections}~\cite{allcott2017social, Bovet2019}, for example. 

In recent years, various applications and tools for misinformation detection were presented, such as \textit{CredEye}~\cite{Popat2018}, \textit{Grover}~\cite{Zellers2019}, \textit{BRENDA}~\cite{Botnevik2020}, and \textit{FactCatch}~\cite{Nguyen2020a}. They support users and manual fact-checking efforts like \textit{Snopes} and \textit{Politifact} to evaluate facts and the credibility of claims. 
However, typically different modalities such as image and text are leveraged to convey news and information more efficiently and attract attention. To comprehend the overall multimodal message and meaning, it is crucial to understand the complex interplay of the modalities. Thus, it is becoming an increasingly important task to develop automated tools and systems for information extraction in multimedia content in order to, e.g., evaluate the overall multimodal message, facilitate semantic search, or analyze the content for credibility.

So far, only a few approaches have explored mutual information and semantic correlations~\cite{Henning2017, Zhang2018, Ye2019, Kruk2019, Otto2019} between both modalities to bridge the semantic gap~\cite{Smeulders2000} or intended to reveal inconsistencies between image-text-pairs with respect to entity representations~(\textit{persons}, \textit{locations}, \textit{organizations}, etc.) to identify repurposed multimedia content~\cite{Jaiswal2017, Sabir2018, Jaiswal2019, MuellerBudack2020}. Unfortunately, these research approaches have not introduced publicly available services that are easily accessible to a broad audience. 

In this paper, we present a Web application called \textit{\appname{}}\footnote{\url{https://labs.tib.eu/newsanalytics}} that uses state-of-the-art approaches from both \textit{Natural Language Processing} and \textit{Computer Vision} to quantify the cross-modal consistency across image and text of the entity types \textit{persons}, \textit{location}, and \textit{events}. For each named entity detected in the text, users are automatically provided with measures of cross-modal similarity, knowledge base information, and exemplary images to evaluate image-text relations in news, tweets, and other multimedia content. 
%

\textit{\appname{}} can be used by a broad audience for applications in various domains due to its simplicity and unsupervised design that can cope with the dynamic nature of news and social media topics: 
(1)~Human assessors and fact-checking efforts such as \textit{Snopes} and \textit{Politifact} can use the demonstrator to check and evaluate the credibility of entity mentions in news and tweets. 
(2)~Linguists, semioticians, and communication scientists can conduct empirical studies on cross-modal relations to investigate the usage of multimodal information in news and other documents. 
(3)~The displayed information consisting of linked knowledge graph data and exemplary images of the entities found in the text provide additional context. This additional information might help users understand certain aspects of a particular news topic and explore multimedia articles more efficiently.

\section{Cross-modal Entity Relations of Image and Text}
\label{sec:methodology}
The proposed Web application for the quantification of cross-modal entity relations is based on own previous work~\cite{MuellerBudack2020, MuellerBudack2021a}. Named entity linking~(Section~\ref{sec:nel}) is applied to extract persons, locations, and events from the text. Subsequently, reference images are crawled automatically from the Web and compared to the news image using suitable feature descriptors for the different entity types as explained in Section~\ref{sec:cms}. The expected performance for different entity types is briefly discussed in Section~\ref{sec:results}.
For more details, we refer to M\"uller-Budack et al.~\cite{MuellerBudack2020, MuellerBudack2021a}.

\subsection{Named Entity Linking}
\label{sec:nel}
To verify image-text relations of \textit{persons}, \textit{locations}, and \textit{events}, we apply named entity recognition and disambiguation to link named entities in the text to \textit{Wikipedia} and \textit{Wikidata}. 
%
For each named entity recognized by \textit{spaCy}~\cite{honnibal2017spacy} we select the linked entity candidate with the highest \textit{PageRank} according to \textit{Wikifier}~\cite{Brank2018} for the corresponding text span. In case \textit{spaCy} recognizes an entity that has not been covered by \textit{Wikifier}, the \textit{Wikidata}~API function \textit{"wbsearchentities"} is used for disambiguation. 

The entity type of the linked entities is obtained based on the associated \textit{Wikidata} information. 
Entities that are an instance~(\textit{Wikidata property P31}) of~\textit{human}~(Wikidata value \textit{Q5}) according to \textit{Wikidata} are considered as \textit{persons}, while for \textit{locations}~(e.g., cities, landmarks, and buildings) a valid \textit{coordinate location}~(\textit{Wikidata property P625}) is set as a requirement. A named entity is considered as an \textit{event} if it is covered in a verified list of events\footnote{Verified events from \textit{EventKG}: \url{http://eventkg.l3s.uni-hannover.de/data/event\_list.tsv}} according to \textit{EventKG}~\cite{Gottschalk2018, Gottschalk2019}. Linked entities that do not fulfill any of the aforementioned criteria are neglected. 

Currently, the demonstrator supports \textit{English} and \textit{German} as languages. However, other languages can be easily integrated and will be supported in the future.
\subsection{Quantification of Cross-modal Similarities}
\label{sec:cms}
Based on the entities detected by the named entity linking approach, the goal is to quantify the cross-modal consistency in image and text. Since we believe that supervised approaches for misinformation detection can not generalize well to future topics and entities that evolve in news~(unless with frequent updates and re-training), the proposed system is completely unsupervised. 
For each named entity detected in the text, visual evidence, i.e., reference images are crawled automatically from the Web. Several image search engines such as \textit{Bing} or \textit{Google} are available. In this demo, a maximum number of $k=5$~reference images is crawled from \textit{Bing} for each entity using its Wikidata label as the search query. This allows us to compare visual features extracted by suitable deep learning approaches of the reference images to the query (news) image. 

\paragraph{Feature Extraction:}
For person entities, an implementation\footnote{Implementation from \textit{David Sandberg}: \url{https://github.com/davidsandberg/facenet}} of \textit{FaceNet}~\cite{Schroff2015} is used to calculate facial features for each face detected by the \textit{Multi-task Cascaded Convolutional Networks}~\cite{Zhang2016a}. An agglomerative clustering approach is applied to calculate the mean feature vector of the majority cluster that most likely represents a queried person.
Geographical features for location entities are extracted using the \textit{base}\,($M,f\text{*}$) model\footnote{\textit{base}\,($M,f\text{*}$) geolocalization model: \url{https://github.com/TIBHannover/GeoEstimation}} for geolocation estimation~\cite{MuellerBudack2018a}. 
Due to the absence of suitable models for event classification, our previous solution \citet{MuellerBudack2020} has used a more general model for scene/place classification to extract features for event verification. Recently, we have presented an event classification approach~\cite{MuellerBudack2021} that achieves better results for this task~\cite{MuellerBudack2021a}. Thus, the proposed application uses the ontology-driven $CO^{cos}_{\gamma}$~model\footnote{$CO^{cos}_{\gamma}$ event classification model: \url{https://github.com/TIBHannover/VisE}} to extract features for event entities.

\paragraph{Measures of Cross-modal Similarity:}
Ultimately, the extracted feature vector(s) of the article's image is compared to all feature vector(s) extracted from the reference images of a named entity using the cosine similarity. The maximum similarity among those comparisons defines the cross-modal similarity~(CMPS for persons, CMLS for locations, and CMES for events) of this particular entity. Note, that the amount of facial feature vectors extracted from an image depends on the number of detected faces.  

\subsection{Quantitative Results}
\label{sec:results}
The performance of the proposed solution was evaluated in previous work~\cite{MuellerBudack2020, MuellerBudack2021a} for two tasks, namely document verification and collection retrieval. Results for the revised \textit{TamperedNews} dataset~\cite{MuellerBudack2021a} are reported in Table~\ref{tab:cms_results_tamperednews}. 
\renewcommand{\b}{\textbf}
\begin{table}[t]
\caption{Verification Accuracy~(VA) and Area Under Receiver Operating Curve~(AUC) for the \textit{TamperedNews~(Top-50\%)} dataset for different entity test sets.}
\label{tab:cms_results_tamperednews}
\fontsize{8}{8}\selectfont
\setlength\tabcolsep{4pt}
\renewcommand{\arraystretch}{1.4}
\centering
    \begin{tabularx}{\linewidth}{X | c | c }

        \toprule
        
        
        \b{Tampering Strategy (Test Set)} & \b{VA} & \b{AUC} \\
        
        \midrule
        \multicolumn{3}{c}{\b{Persons} (\num{16848} documents)}\\
        \midrule
        Random person                                   & 0.94 & 0.95 \\ 
        ... with same Country of Citizenship (PsC)      & 0.93 & 0.94 \\ 
        ... with same Gender (PsG)                      & 0.94 & 0.95 \\ 
        ... with same Country \& Gender (PsCG)          & 0.93 & 0.94 \\ 
        
        \midrule
        
        \multicolumn{3}{c}{\b{Locations - Outdoor} (\num{14113} documents)}\\
        \midrule
        Random location                                                         & 0.88 & 0.85 \\ 
        ... of same type and distance from 750 - 2500 km (GCD$_{200}^{750}$)    & 0.86 & 0.81 \\ 
        ... of same type and distance from 200 - 750 km (GCD$_{200}^{750}$)     & 0.79 & 0.74 \\ 
        ... of same type and distance from 25 - 200 km (GCD$_{25}^{200}$)       & 0.77 & 0.72 \\ 

        \midrule
        \multicolumn{3}{c}{\b{Locations - Indoor} (\num{19129} documents)}\\
        \midrule
        Random location                                                         & 0.74 & 0.72 \\ 
        ... of same type and distance from 750 - 2500 km (GCD$_{200}^{750}$)    & 0.73 & 0.70 \\ 
        ... of same type and distance from 200 - 750 km (GCD$_{200}^{750}$)     & 0.74 & 0.71 \\ 
        ... of same type and distance from 25 - 200 km (GCD$_{25}^{200}$)       & 0.69 & 0.68 \\ 
        
        \midrule
        
        \multicolumn{3}{c}{\b{Events} (\num{7734} documents)}\\
        \midrule
        Random event                                                & 0.92 & 0.91 \\ 
        ... with same parent class in \textit{Wikidata} (EsP)       & 0.75 & 0.71 \\ 
        
        
        \bottomrule
        
    \end{tabularx}
\end{table}

For evaluation, we have tampered the articles by replacing each original~(untampered) entity with another random entity or an entity that shares certain attributes for more challenging test sets. 
The verification accuracy~(VA) quantifies how many of the test documents of the untampered entity set have achieved a higher cross-modal similarity compared to \textit{one} tampered entity set for the respective document's image. 
For collection retrieval, all $|\set{D}|$~untampered documents in the dataset as well as the respective $|\set{D}|$~tampered documents of \textit{one} tampering strategy are considered. \textit{Area Under Receiver Operating Curve}~(AUC) is the evaluation metric.

Overall, promising results have been achieved. In particular, the performance for person and outdoor location verification implies that the system can reliably quantify relations for entities of these types. Indoor locations are more challenging to verify since they typically contain few geographical cues that can also be ambiguous~\cite{MuellerBudack2018a}. Results for event verification are worse when tampering with events that share a parent class~(e.g., tampering the \textit{2016 US elections} with the \textit{2020 US election}) as they depict very similar scenes. In general, it is more challenging to verify locations and events compared to persons since reference images do not necessarily match the news content. For example, querying images for a "coarse" location entity such as a continent very rarely depict relevant imagery for verification. A detailed evaluation for particular types of locations and events is conducted in previous work~\cite{MuellerBudack2020, MuellerBudack2021a}.  
\section{Implementation}
\label{sec:implementation}
\begin{figure}[t]
	\centering
    \includegraphics[width=1\linewidth]{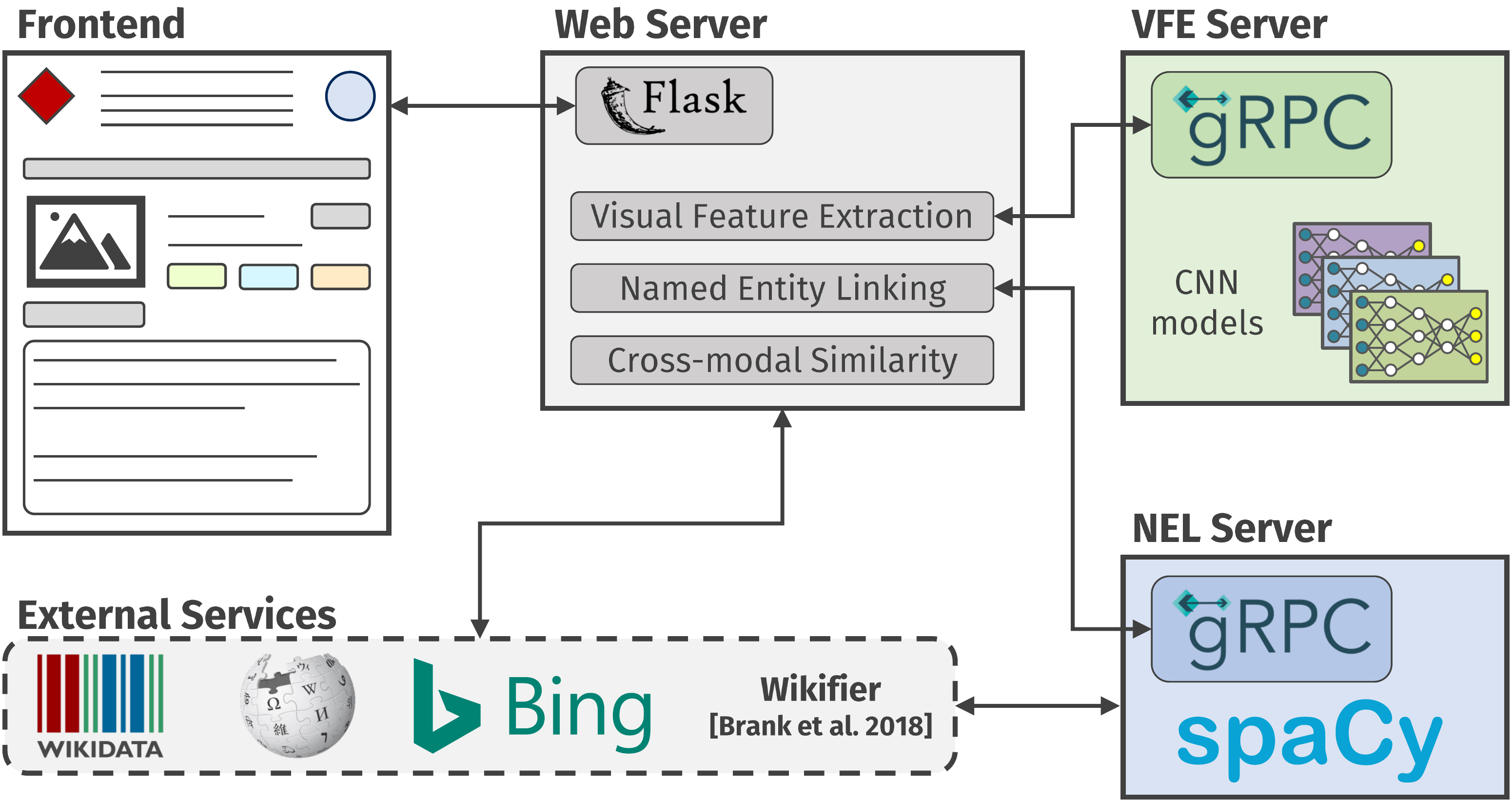}
	\caption{Overview of the proposed demonstrator. A REST API establishes the communication between frontend and Web server. The Web server uses Google's Remote Procedure Calls~(gRPC) to communicate with the respective services for named entity linking~(NEL, Section~\ref{sec:nel}) and visual feature extraction~(VFE, Section~\ref{sec:cms}).} 
	\label{fig:technical_overview}
\end{figure}

The application was split into several microservices to enable their independent development, replacement, and execution. In this way, it is possible to deploy the visual feature extraction~(VFE) server that uses deep learning approaches on a machine with graphic processing units~(GPU) to optimize the runtime performance. In the following, we outline the details of the individual microservices. An overview is presented in Figure~\ref{fig:technical_overview}.

\paragraph{Frontend:} The user interface was implemented as a single page application using the \textit{Vue.js} framework as it allows for a modern and responsive Web design. To ensure that the interface works on as many devices and resolutions as possible, \textit{Bootstrap} was used for the design and layout of the application.

\paragraph{Web Server:} The Web server is implemented as a \textit{Flask} server with \textit{REST API} using \textit{Python} and serves as a communication hub between the user interface, the two \textit{gRPC} (Google's Remote Procedure Calls) services for named entity linking~(NEL) and visual feature extraction~(VFE), and the external resources.
%
%
To prevent timeouts that can be caused by texts with excessive length and thus a large number of named entities, we adopt \textit{Celery} as an asynchronous task scheduler. During the computation, the User is constantly provided with the current computation status for a more responsive feel. Details on runtimes can be found at the end of this section. To increase performance and reduce the number of calls to external sources, the individual requests are cached for 24 hours.

\paragraph{NEL Server:} Based on the user's input, the Web server forwards the input text to the named entity linking~(NEL) server using \textit{gRPC}. Then, the NEL server performs the approach described in Section~\ref{sec:nel} and returns the spans of all named entities including its linked \textit{Wikidata} and \textit{Wikipedia} information.

\paragraph{VFE Server:} The visual feature extraction~(VFE) server was designed as a \textit{gRPC} service and is deployed on a machine that uses a single \textit{NVIDIA GeForce Titan X} graphics card. It uses \textit{PyTorch} or \textit{TensorFlow} depending on the implementation of the respective deep learning model applied for face~\cite{Schroff2015}, location~\cite{MuellerBudack2018a}, and event entities~\cite{MuellerBudack2021}. The images are transferred from the Web server to the VFE Server as an encoded binary string. The feature vectors are returned to the Web server which subsequently calculated the cross-modal entity similarities.

\paragraph{Runtime:} 
The proposed application depends on the response times of \textit{Bing}, \textit{Wikidata}, \textit{Wikipedia}, and \textit{Wikifier}. As a result, the runtime is difficult to specify. However, for regular news articles, computations for individual steps rarely exceed ten seconds. Longest response times were noticed for person verification as an additional face detection step is required. This was tested for five articles~(cached was emptied before) with an average of approximately \num{4000} characters~(from 1500 to 6000), 6 unique persons~(from 3 to 12), 3 unique locations~(from 1 to 7), and 4 unique events~(from 0 to 8).  Given the complexity of the approach and the integrated cache, we believe that the computation times are acceptable.
\section{Demonstrator}
\label{sec:demo}
\begin{figure*}[t]
	\centering
    \includegraphics[width=1\linewidth]{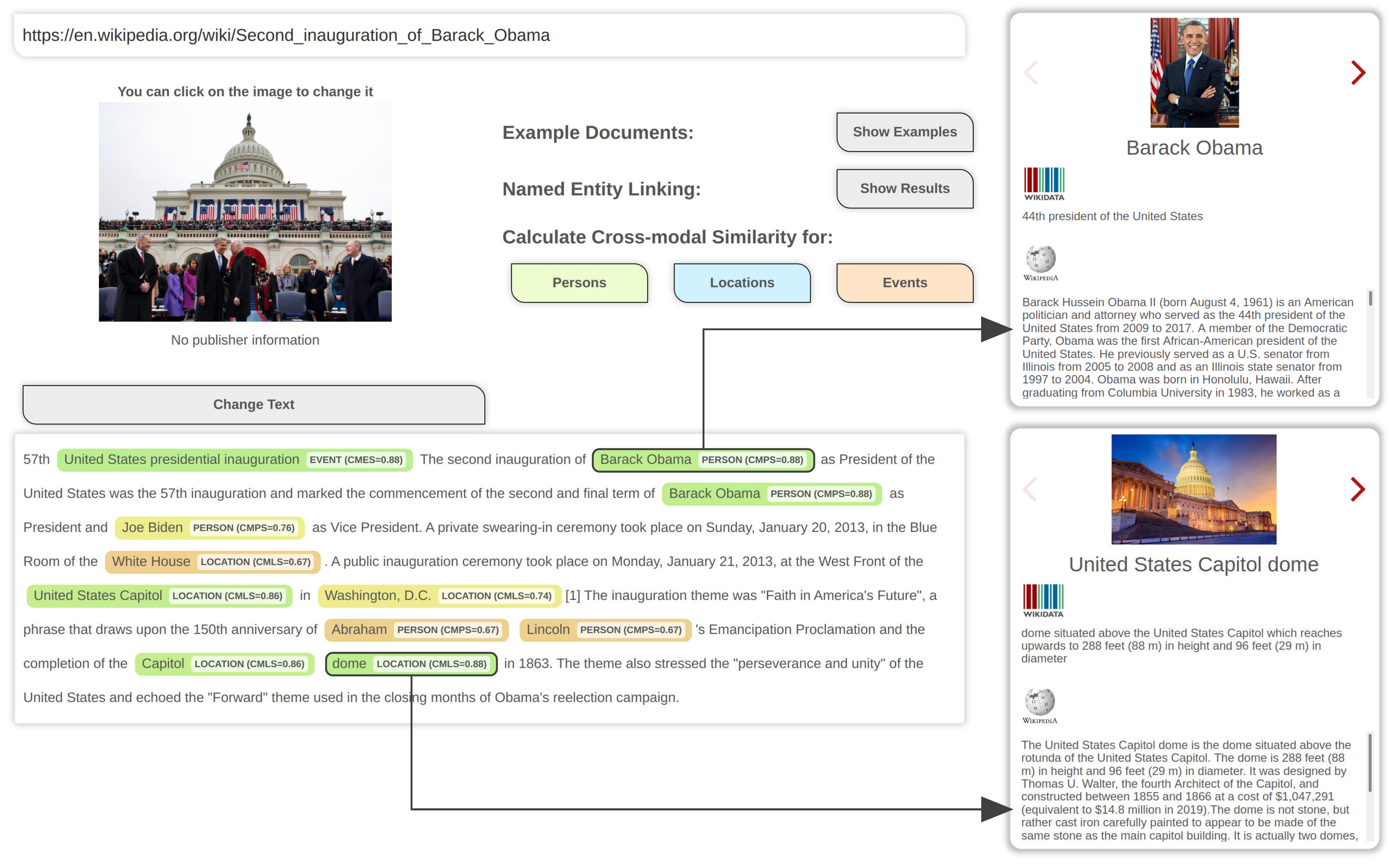}
	\caption{Screenshot of the demonstrator for multimodal news analytics. Left: Exemplary result for the quantification of cross-modal similarities in a multimodal article. Right: The user can hover over specific entities to obtain further entity information from \textit{Wikipedia} and \textit{Wikidata} as well as to explore the respective reference images.}
	\label{fig:cms_demo}
\end{figure*}

The demonstrator of the proposed system is publicly available at \url{https://labs.tib.eu/newsanalytics}. The source code will be provided on our GitHub page\footnote{\url{https://github.com/TIBHannover/cross-modal_entity_consistency}}. Screenshots depicting the functionality are presented in Figure~\ref{fig:cms_demo}. 
The demonstrator allows users to either paste a Web link of a multimodal (news) article or to alternatively upload a self-defined image along with a text. In this way, it is possible to analyze arbitrary image-text pairs ranging from long news documents over tweets to scientific articles. Moreover, the system can be applied to metadata verification by defining a mention of a single claimed entity as "text" to specifically determine its visual probability in the given image. If a Web link is provided, the text and main~(top) image are parsed from the Web link using the \textit{newspaper3k} Python library and subsequently shown to the user.

Once an image-text pair has been selected or the text has been modified by the user, the named entity linking is automatically applied as explained in Section~\ref{sec:nel} to extract mentions of \textit{persons}, \textit{locations}, and \textit{events}. By hovering over specific entities, an information card presents the user an overview including entity descriptions extracted from \textit{Wikipedia} and \textit{Wikidata} as well as an image, if available, using the \textit{Wikidata} property~\textit{P18}. In this context, the user can directly go to the respective Wikidata or Wikipedia page of the entity for further information by clicking on the respective symbol. For some applications, these system outputs might be already sufficient to quickly inform users over specific news topics. 

Finally, the user can click on the corresponding buttons to let the system calculate the proposed \textit{Cross-modal Similarity} for the detected \textit{persons}, \textit{locations}, and \textit{events} to the specified image according to Section~\ref{sec:cms}. In this step, the automatically crawled Web images for the named entities can be explored by hovering over the corresponding entities in the text. Although the user is automatically provided with measures of cross-modal entity similarity, the overview containing additional entity information from \textit{Wikipedia} and \textit{Wikidata}, as well as exemplary images, also allows for a more reliable manual assessment of cross-modal relation in news, tweets, and other multimedia articles.
%



%
\section{Conclusions}
\label{sec:conc}
In this paper, we have presented a novel Web application called \textit{\appname{}} that automatically quantifies relations between image and text in news, tweets, and other multimodal articles. The system utilizes suitable Natural Language Processing and Computer Vision approaches to compute measures of cross-modal entity consistency, which can be helpful, for instance, for manual fact-checking efforts. Additional information from knowledge bases and reference images for the entities detected in the text can further support human assessors to understand the overall multimodal message and meaning as well as to evaluate the credibility of an article. 
In the future, we plan to integrate additional languages into our system. Furthermore, it would be interesting to analyze image-text-relations of additional entity types such as organization or times~(dates). Finally, we plan to investigate recent deep learning approaches for named entity linking that could improve the system's performance. 

%
\bibliographystyle{ACM-Reference-Format}
\bibliography{main}

\end{document}